\documentclass[aps,prl,twocolumn,amssymb,showpacs,superscriptaddress]{revtex4-1}
\usepackage{graphicx}

\newcommand{\ecm}{\={e}/cm$^2$}

\begin{document}

\title{ Comparative study of the effects of electron irradiation and\ natural disorder in single\ crystals of SrFe$_{2}$(As$_{1-x}$P$_x$)$_2$ ($x=$0.35) superconductor}

\author{C.~P.~Strehlow}
\affiliation{The Ames Laboratory and Department of Physics \&\ Astronomy, Iowa State University, Ames, Iowa 50011, USA}

\author{M. Ko\'{n}czykowski}
\affiliation{Laboratoire des Solides Irradi\'{e}s, CNRS-UMR 7642 \& CEA-DSM-IRAMIS, Ecole Polytechnique, F91128 Palaiseau cedex, France}

\author{J.~A.~Murphy}
\affiliation{The Ames Laboratory and Department of Physics \&\ Astronomy, Iowa State University, Ames, Iowa 50011, USA}

\author{S. Teknowijoyo}
\affiliation{The Ames Laboratory and Department of Physics \&\ Astronomy, Iowa State University, Ames, Iowa 50011, USA}

\author{K.~Cho}
\affiliation{The Ames Laboratory and Department of Physics \&\ Astronomy, Iowa State University, Ames, Iowa 50011, USA}

\author{M.~A.~Tanatar}
\affiliation{The Ames Laboratory and Department of Physics \&\ Astronomy, Iowa State University, Ames, Iowa 50011, USA}

\author{T.~Kobayashi}
\affiliation{Department of Physics, Graduate School of Science, Osaka University, Toyonaka, Osaka 560-0043, Japan}

\author{S.~Miyasaka}
\affiliation{Department of Physics, Graduate School of Science, Osaka University, Toyonaka, Osaka 560-0043, Japan}
\affiliation{JST, Transformative Research-Project on Iron-Pnictides (TRIP), Chiyoda, Tokyo 102-0075, Japan}

\author{S.~Tajima}
\affiliation{Department of Physics, Graduate School of Science, Osaka University, Toyonaka, Osaka 560-0043, Japan}
\affiliation{JST, Transformative Research-Project on Iron-Pnictides (TRIP), Chiyoda, Tokyo 102-0075, Japan}

\author{R.~Prozorov}
\email[corresponding author: ]{prozorov@ameslab.gov}
\affiliation{The Ames Laboratory and Department of Physics \&\ Astronomy, Iowa State University, Ames, Iowa 50011, USA}

\date{27 May 2014}

\begin{abstract}

London penetration depth, $\lambda(T)$, was measured in single crystals of SrFe$_2$(As$_{1-x}$P$_x$)$_2$ ($x=$0.35) iron - based superconductor. The influence of disorder on the transition temperature, $T_c$, and on $\lambda(T)$ was investigated. The effects of scattering controlled by the annealing of as-grown crystals was compared with the effects of artificial disorder introduced by  2.5~MeV electron irradiation. The low temperature behavior of $\lambda(T)$ can be described by a power-law function, $\Delta \lambda (T)=AT^n$, with the exponent $n$ close to one in pristine annealed samples, as expected for superconducting gap with line nodes. Upon $1.2 \times 10^{19}$ \ecm irradiation, the exponent $n$ increases rapidly exceeding a dirty limit value of $n=$ 2 implying that the nodes in the superconducting gap are accidental and can be lifted by the disorder. The variation of the exponent $n$ with  $T_c$ is much stronger in the irradiated
  crystals compared to the crystals in which disorder was controlled by the annealing of the growth defects.
We discuss the results in terms of different influence of different types of disorder on intra- and inter- band scattering.
\end{abstract}



\maketitle

The pairing mechanism in Fe-based superconductors has been the focal point of many theoretical and experimental works \cite{Chubukov-review,Hirschfeld-review}. Proximity to magnetism and  high superconducting transition temperatures, $T_c$, in upper 50~K range, prompted the search for non phonon mechanism of superconductivity \cite{phonon}. Based on early experiments Mazin {\it et al.} \cite{Mazinspm1,MazinvNature} suggested unconventional superconducting state with interband pairing in which superconducting gap function changes sign between different sheets of the Fermi surface, but remains full (without line nodes) on each sheet. Experimental verification of this so called $s_\pm$ pairing mechanism quickly became a key point of superconducting gap structure studies in iron-based materials.

The verification of the $k$-space sign-changing gap in iron-based superconductors turned out to be more difficult than it was in the cuprates, in which sign change along a single Fermi surface was proven by directional phase sensitive experiments \cite{vanHarlingenRMP1995,TsueiKirtley}. It was suggested that impurity scattering can be used as a probe of a sign-changing gap \cite{impurities}. This approach becomes significantly more powerful when in addition to the suppression of $T_c$, other thermodynamic quantities, such as London penetration depth, are studied on the same samples  \cite{PH,Hyunsooheavyion}.

Artificial disorder in superconductors can be introduced in a controlled way by irradiation. Depending on the irradiation type and energy, the induced defects have different characteristics. Early studies of $T_c$ \cite{columnar,Hyunsooheavyion,Jason} and $\lambda (T)$ in Ba(Fe$_{1-x}$Co$_x$)$_2$As$_2$ (BaCo122) \cite{Hyunsooheavyion} and Ba(Fe$_{1-x}$Ni$_x$)$_2$As$_2$ (BaNi122) \cite{Hyunsooheavyion,Jason} used heavy ion irradiation that is known to create one-dimensional columnar defects \cite{columnar}. The analysis of $T_c$ and of the exponent $n$ of the power-law function used to fit temperature-dependent London penetration depth, $\Delta \lambda=AT^n$, was consistent with the predictions of $s_\pm$ model. Similar study in the optimally hole-doped (Ba$_{1-x}$K$_x$)Fe$_2$As$_2$ ($x=$0.4) \cite{Salovich} found exponential to $\Delta \lambda (T) \sim T^2$ crossover at very high irradiation doses, but virtually no change in $T_c$ most probably indicating dominant intra-band pairing interaction in this compound \cite{PH}.  Irradiation with 2 MeV $\alpha$-particles \cite{alpha} and 3 MeV protons \cite{protons1,protons2,protons3}, both creating cluster-like defects \cite{Damask1963}, found much faster suppression of $T_c$ than in the case of columnar defects, but still much slower than originally predicted for a simplified ``symmetric" $s_\pm$ scenario \cite{Kontani2009}. More recently, the predictions for the $s_\pm$ scenario  were significantly relaxed in a realistic ``asymmetric" model \cite{PH}. These calculations were used to fit a significant variation of $T_c$ induced by 2.5 MeV electron irradiation in BaRu122 \cite{electrons2}. Similar suppression rate was found in other 122 compounds, including BaCo122 and Ba(AsP)122 \cite{electrons1}.
In the material of this study, SrFe$_2$(As$_{1-x}$P$_x$)$_2$, the effect of post-growth disorder was studied previously by measuring both $T_c$ and $\lambda (T)$  in samples before and after annealing \cite{JasonSrP}. It was known that annealing annealing of  optimally ($x=$ 0.35) substituted samples leads to enhancement of $T_c$ from typical 25 to 27~K to almost 35~K \cite{SrP122annealing,Kobayashi2013}. The analysis of the low-temperature behavior of the penetration depth was unambiguously consistent with the presence of line nodes in the gap \cite{JasonSrP}, very similar to another material with isovalent P substitution, BaFe$_2$(As$_{1-x}$P$_x$)$_2$ \cite{AsP122Science2012}.

In this paper we report a comparative study of the effects of artificial and natural disorder on $T_c$ and quasi-particle excitations/superconducting gap structure in single crystals of SrFe$_2$(As$_{1-x}$P$_x$)$_2$ with optimal level of isovalent phosphorus substitution, $x=$0.35. Our main observation is that electron irradiation and natural defects change $T_c$ and London penetration depth, $\lambda (T)$,  $\lambda (T)$, in significantly distinct ways. We relate this dissimilarity to a possible difference in the scattering amplitude and characteristic spatial range of the scattering potential, which may also include residual after-growth strain.

Single crystals of SrFe$_2$(As$_{1-x}$P$_{x}$)$_2$ with optimal isovalent substitution level, $x$=0.35, were grown from stoichiometric mixtures of Sr, FeAs, and FeP powders as described in detail in Refs.~\cite{SrP122annealing,Kobayashi2013}.
The samples studied were from two different batches, $A$ and $B$. They were cleaved with a razor blade from the inner parts of larger single crystals and had two shiny cleavage surfaces, and thickness of about 60 to 70 $\mu$m. Side surfaces of the samples were also cleaved along $a$-axes in the plane, and the samples were close to 0.6 $\times$ 0.6 mm$^2$ in the surface area. Prior to penetration depth measurements, the same samples were measured using magneto-optical technique \cite{MO} to check for possible cracks and macroscopic inhomogeneity. We found typical Bean profile of the magnetic induction distribution with no visible anomalies, reflecting high sample quality and good magnetic uniformity.

Measurements of in-plane London penetration depth were performed by using a self - resonating tunnel diode resonator (TDR), which is essentially a radio-frequency (14 MHz) magnetic susceptibility measurement \cite{Prozorov1,Prozorov2}.
The 2.5 MeV electron irradiation was performed at the SIRIUS Pelletron linear accelerator operated by the \textit{Laboratoire des Solides Irradi\'{e}s} (LSI) at the \textit{Ecole Polytechnique} in Palaiseau, France. The irradiation dose is represented here in C/cm$^2$. To convert to electrons per cm$^2$, this number needs to be divided by the electron charge, \={e}. The sample of batch $A$ was exposed to a dose of 2.2 C/cm$^2$ and the sample of batch $B$ was exposed to 1.1 C/cm$^2$. After the irradiation the samples were warmed up to a room temperature, which results in up to 30\% partial annealing of the defects \cite{electrons2}. Importantly, the comparative measurements of the effect of irradiation and of the annealing were conducted on physically the same samples before and after treatment.


\begin{figure}[tb]
\begin{centering}
\includegraphics[width=0.9\linewidth]{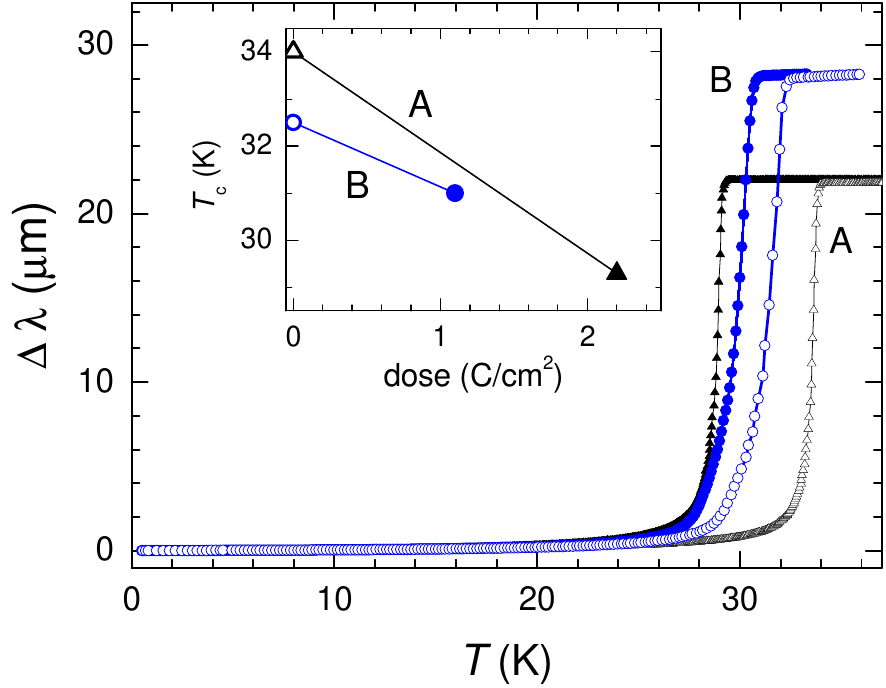}
\par\end{centering}
\caption{(Color Online) Full temperature range variation of the London penetration depth, $\Delta \lambda(T)$, in two single crystals of SrFe$_2$(As$_{1-x}$P$_{x}$)$_2$, $x$=0.35, $A$ (black triangles) and $B$ (blue circles) before (open symbols) and after (solid symbols) irradiation with doses of 2.2 C/cm$^2$ and 1.1 C/cm$^2$, respectively. Inset shows the change of $T_c$ as a function of the irradiation dose. }
\label{Figure1}
\end{figure}

\begin{figure}[tb]
\begin{centering}
\includegraphics[width=1.0\linewidth]{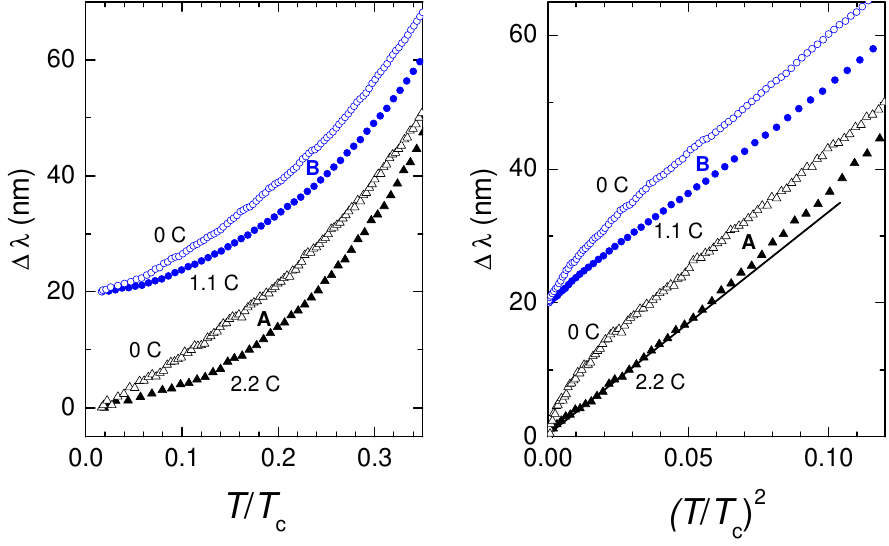}
\par\end{centering}
\caption{(Color Online) Low-temperature variation of the London penetration depth, $\Delta \lambda(T)$, in single crystals of SrFe$_2$(As$_{1-x}$P$_{x}$)$_2$, $x$=0.35, plotted vs. reduced temperature $T/T_c$ (left panel) and vs. $(T/T_c)^2$ (right panel). The data for sample $A$ in the pristine annealed state are shown by open black triangles, and after the irradiation with 2.2 C/cm$^2$ by solid black triangles. The data for sample $B$ before (blue open circles) and after 1.1 C/cm$^2$ irradiation (blue solid circles) are offset by 20 nm to avoid overlapping. Line in the right panel is guide to the eye to show a slight upward curvature suggesting $n >2$ after electron irradiation with 2.2 C/cm$^2$.
}
\label{Figure2}
\end{figure}

Figure~\ref{Figure1} shows the variation of the London penetration depth in samples $A$ and $B$ over the whole temperature range from the base temperature of 0.5~K to above $T_c$. Empty and closed symbols show the data for the same sample before and after irradiation. Inset in Fig.~\ref{Figure1} shows that two studied samples exhibit similar slope of $T_c$ vs. the irradiation dose. Also, both samples show sharp superconducting transitions before and after the irradiation, suggesting spatially homogeneous distribution of the induced defects. This is not strange considering that the electrons at energy of 2.5 MeV have stoppage distance of more than 500 $\mu$m, significantly longer than sample thickness.

\begin{figure}[tb]
\begin{centering}
\includegraphics[width=1.0\linewidth]{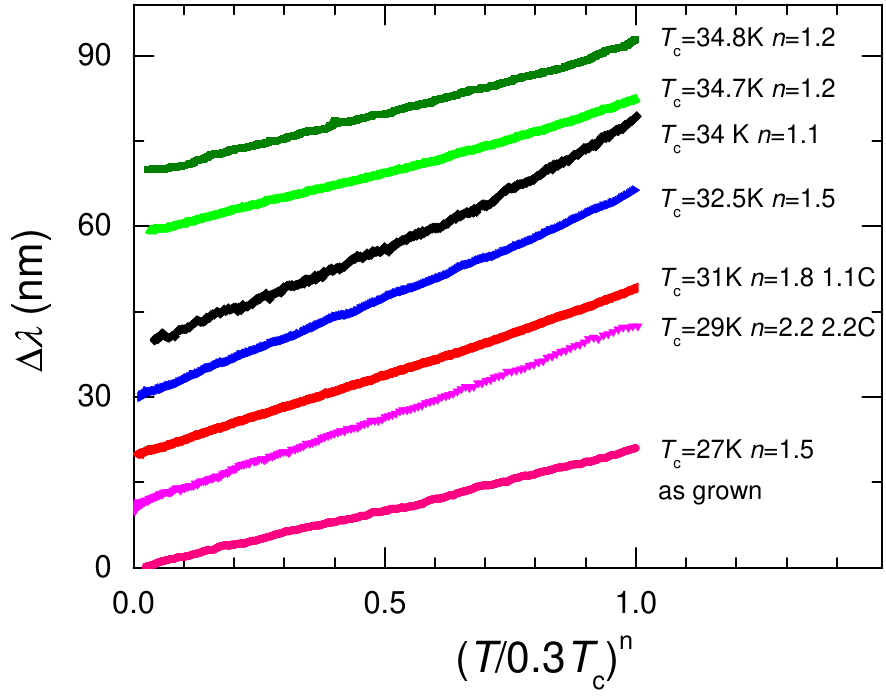}
\par\end{centering}
\caption{(Color Online) $\Delta \lambda (T)$, plotted vs. reduced temperature $(T/0.3T_c)^n$, with $n$ as a fitting parameter selected to linearize the data. The data for pristine annealed samples (top four curves, $T_c \geq $32.5 K) are well described with a singe power law each over the whole range. The data for samples with growth defects show $n=$1.5, while the samples subjected to electron irradiation show a rapid increase of $n$ with $T_c$ suppression. Note that for all samples $\Delta \lambda (0.3T_c$ is about the same.}
\label{Figure3}
\end{figure}

Figure \ref{Figure2} zooms at the low-temperature variation of the London penetration depth, $\Delta\lambda (T)$, in pristine and irradiated samples $A$ and $B$, up to $T_c/3$. In this regime, the temperature dependence of the superconducting gap becomes negligible and the total variation is determined by the temperature - induced population of quasiparticles. In a clean case this leads to the exponential $\Delta \lambda (T)$ in full gap superconductors and to $T-$linear behavior in case of a gap with line nodes. In the dirty limit, both converge to a $T^2$ dependence. The usual way of analysis is to fit the experimental $\lambda(T)$ to a power-law function, $\Delta \lambda (T)=AT^n$. In Fig.~\ref{Figure2} we plot the penetration depth for two samples vs. the reduced temperature, $T/T_c$, (left panel) and vs. $(T/T_c)^2$ (right panel). The data for samples in pristine state before irradiation are in reasonable agreement with the results of o
 ur previous study, see Fig.~\ref{Figure3} below \cite{JasonSrP}. The electron irradiation significantly decreases total variation of $\Delta \lambda (T)$ reflected in an increased exponent $n$. In sample $B$ subjected to 1.1 C/cm$^2$ the exponent $n=$1.8. Direct fitting of the data for sample $A$, exposed to 2.2 C/cm$^2$ irradiation results in $n=$2.26. Plotting the data vs. $(T/T_c)^2$, right panel of Fig.~\ref{Figure2}, shows that sample with larger dose reveals an upward curvature, suggesting that $n>2$.

\begin{figure}[tb]
\begin{centering}
\includegraphics[width=0.9\linewidth]{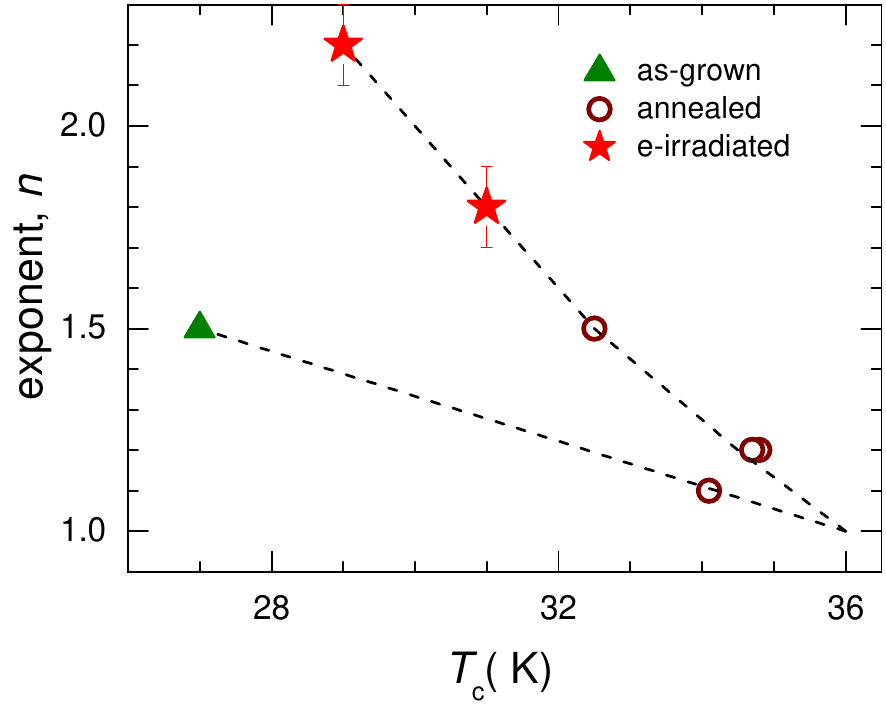}
\par\end{centering}
\caption{(Color Online) The exponent $n$ of the power-law fit of $\Delta \lambda (T)$ from Fig.~\ref{Figure3}, vs. $T_c$. Note significantly smaller exponents for as grown and annealed compared to the samples with irradiation defects.
}
\label{Figure4}
\end{figure}

It is generally accepted that the exponent $n>$2 cannot be explained by the effect of disorder in superconductors with symmetry imposed line nodes. There is, however, a caveat, that in a multi-band superconductor the range over which a characteristic $T^2$ dependence is observed, can be significantly smaller that in a single gap superconductor. To study this possibility, in Fig.~\ref{Figure3} we performed a different way of the analysis of the functional form of $\Delta \lambda (T)$. Here $\Delta \lambda (T)$ for each sample was plotted vs. $(T/0.3T_c)^n$, where $n$ was chosen to produce the closest to linear dependence. Previously, we showed that the data for both as-grown low $T_c$ samples and annealed high $T_c$ samples can be well described by using Hirschfeld-Goldenfeld interpolation formula \cite{HG,JasonSrP}, with the effective temperature $T^*$ increasing with the amount of disorder. The data for these samples can be actually linearized using an exponent $n$ close to 1, for all samples
  with $T_c>$34~K clearly suggesting line nodes in the superconducting gap. The exponent $n$ increases for samples with lower $T_c$. However, this linearization procedure does not produce non-monotonic dependencies as expected for pronounced multi-gap effects. Therefore our data are best described by a true power law with $n>2$ in the most irradiated sample.

To summarize our findings, Figure~\ref{Figure4} shows the exponent $n$ plotted vs. $T_c$, controlled either by the growth defects or by the defects induced by electron irradiation. The data for the two types of disorder reveal striking dissimilarity. While the variation of both $n$ and $T_c$ in the samples with growth disorder is consistent with impurity effect in superconductors with symmetry - imposed line nodes, irradiation brings the exponent $n$ above the range allowed for such superconductors, despite significantly milder suppression of $T_c$.
Furthermore,  we do not observe any increase of $\Delta \lambda (T)$ on cooling \cite{MartinNd} which could be suggestive of paramagnetic effects induced by the irradiation, so the difference between the effects is unlikely to be non-magnetic vs. magnetic scattering. The irradiation with electrons at energies between 1 and 10 MeV is known to create primarily Frenkel pairs of vacancies and interstitial ions \cite{Damask1963}. Interstitials tend to migrate and disappear at surfaces and other sinks in the crystal structure, leaving vacancies as point - like disorder \cite{Damask1963}. On the other hand, disorder in as grown samples mostly appears as dislocations, which have long - range elastic strain field and therefore these two types of
  defects have significantly different characteristics. It is therefore conceivable that scattering on these defects is characterized by a notably different momentum transfer. In other words, the strength and characteristic range and dimensionality of the scattering potentials corresponding to these defects are quite different.

In multi-band iron-based superconductors scattering in inter- and intra- band channels have very different effect on superconductivity \cite{Chubukov-review,Hirschfeld-review,Korshunov,PH}. Scattering with small moment transfer, would scatter electrons only on the same Fermi surface sheets (intra-band scattering), while to scatter between different sheets of the Fermi surface a large momentum transfer is needed. Since point defects have a characteristic size of a unit cell or less, the characteristic wavevector is of the order of $Q \sim 2\pi/a$, a sizable fraction of the Brillouin zone. For this type of disorder we would naturally expect strong contribution to the inter-band scattering channel. By the same logic, extended defects would be characterized by a much smaller $Q$, and contribute mostly to the intra-band scattering.
This and previous studies show that SrFe$_2$(As$_{1-x}$P$_{x}$)$_2$ has superconducting gap with line nodes \cite{JasonSrP,Dulguun,Maeda}. Small $Q$ scattering within each sheet of the Fermi surface in this situation will be very similar to the usual effect of disorder in nodal superconductors, as we observed here for the growth defects. Large $Q$ scattering on point defects, however, will lift the nodes and drive the gap structure towards the full gap, if the nodes are accidental \cite{PH}. Our observation of $n>2$ supports this scenario. Alternative scenario by Korshunov {\it et al.} suggests a transition from $s_\pm$ to a conventional $s^{++}$ state as a function of disorder \cite{Korshunov}.

In conclusion, we found dramatically different effect of artificial and natural as - grown disorder on superconducting transition temperature, $T_c$, and on quasi-particle excitations measured by the variation of London penetration depth in isoelectron substituted SrFe$_{2}$(As$_{1-x}$P$_x$)$_2$. The response to the post-growth disorder is similar to the usual effect of disorder in nodal superconductors. The response to electron irradiation is notably different suggesting the evolution of the superconducting gap structure from nodal to nodeless, expected for accidental nodes. We relate the difference between the two types of disorder to the difference in the characteristic scattering wavevector $Q$, with small $Q$ for post-growth disorder and large $Q$ for point type disorder.

\emph{Post - submission note}: After this paper was submitted, a related work appeared on arxiv where the Authors study the effects of electron irradiation on both $T_c$ and penetration depth in another isovalent substituted compound, BaFe$_{2}$(As$_{1-x}$P$_x$)$_2$ \cite{Mizukami2014eprint}. Their results support the accidental nodes scenario in this class of nodal iron pnictides.

\acknowledgements The work in Ames was supported by the U.S. Department of Energy (DOE), Office of Science, Basic Energy Sciences, Materials Science and Engineering Division. Ames Laboratory is operated for the U.S. DOE by Iowa State University under contract DE-AC02-07CH11358. We thank the SIRIUS team, B. Boizot, V. Metayer, and J. Losco, for running electron irradiation at \textit{Ecole Polytechnique} (supported by EMIR network, proposal 11-11-0121.) Work at Osaka was partly supported by a Grant-in-Aid IRONSEA from the Japan Science and Technology Agency (JST).

\end{document}